\documentclass[epsfig,amsfonts]{mn}

\usepackage{graphicx}
\usepackage{amsfonts}
\usepackage{amssymb}
\usepackage{amsmath}
\usepackage{colors}
\usepackage[normalem]{ulem}
\usepackage[usenames,dvipsnames]{color}

\title[Discrepancies between CFHTLenS cosmic shear and Planck]{
Discrepancies between CFHTLenS cosmic shear \& Planck: new physics or systematic effects?}
\author[Kitching T. D., Verde L., Heavens, A. F., Jimenez R.]{Thomas D. Kitching$^1$\thanks{t.kitching@ucl.ac.uk}, 
Licia Verde$^{2,3,4}$, Alan F. Heavens$^{5}$, Raul Jimenez$^{2,3}$\\
$^1$Mullard Space Science Laboratory, University College London, Holmbury St Mary, Dorking, Surrey RH5 6NT, U.K.\\
$^2$Radcliffe Institute for Advanced Study, Harvard University, Cambridge 02138, USA.\\
$^3$ICREA \& ICC-UB, University of Barcelona, Marti i Franques 1, Barcelona 08028, Spain.\\
$^4$Institute of Theoretical Astrophysics, University of Oslo, Oslo 0315, Norway.\\
$^5$ICIC, Astrophysics, Imperial College, Blackett Laboratory, Prince Consort Road, London SW7 2AZ, U.K.}

\newcommand{\be}{\begin{equation}}  \newcommand{\ee}{\end{equation}}
  \newcommand{\ba}{\begin{eqnarray}}
\newcommand{\ea}{\end{eqnarray}}  
\newcommand{\nn}{\nonumber\\}

\def\gs{\mathrel{\raise1.16pt\hbox{$>$}\kern-7.0pt %
\lower3.06pt\hbox{{$\scriptstyle \sim$}}}}         %
\def\ls{\mathrel{\raise1.16pt\hbox{$<$}\kern-7.0pt %
\lower3.06pt\hbox{{$\scriptstyle \sim$}}}}         %

\newcommand{\apj}{Astrophysical Journal }
\newcommand{\apjl}{Astrophysical Journal Letters}

\newcommand{\mnras}{Mon.Not.Roy.Astron.Soc. }

\newcommand{\prd}{Physical Review D }

\begin{document}

\voffset=-0.25in 
\maketitle

\begin{abstract}
There is currently a discrepancy in the measured value of the amplitude of matter clustering, parameterised using $\sigma_8$, inferred from galaxy weak lensing, and cosmic microwave background data, which could be an indication of new physics, such as massive neutrinos or a modification to the gravity law, or baryon feedback.
In this paper we make the assumption that the cosmological parameters are well determined by Planck, and use weak lensing data to investigate the implications for baryon feedback and massive neutrinos, as well as possible contributions from intrinsic alignments and biases in photometric redshifts. We apply a non-parametric approach to model the baryonic feedback on the dark matter clustering, which is flexible enough to reproduce the OWLS and Illustris simulation results. The statistic we use, 3D cosmic shear, is a method that extracts cosmological information from weak lensing data using a spherical-Bessel function power spectrum approach. We analyse the CFHTLenS weak lensing data and, assuming best fit cosmological parameters from the Planck CMB experiment, find that there is no evidence for baryonic feedback on the dark matter power spectrum, but there is evidence for a bias in the photometric redshifts in the CFHTLenS data, consistent with a completely independent analysis by Choi et al. (2015), based on spectroscopic redshifts; and that these conclusions are robust to assumptions about the intrinsic alignment systematic. We also find an upper limit on the sum of neutrino masses conditional on other $\Lambda$CDM parameters being fixed, of $< 0.28$ eV ($1\sigma$).
\end{abstract}


\section{Introduction}
\label{Introduction}
Weak lensing of galaxy images, the effect where the 
observed shape of galaxies is distorted by the presence of mass 
perturbations along the line of sight, is a powerful
probe of the matter distribution in the Universe. This is because 
the distortion - 
a change in the third eccentricity, or third flattening (known as `ellipticity'), 
and size of galaxy images - depends on perturbations in the total matter density which, 
because we live in an apparently dark matter-dominated Universe,  
is in principle sensitive to the dark matter power spectrum directly. 
Accessing the matter power spectrum through weak lensing measurements 
results in a statistic that contains a wealth of cosmological information, where observations 
as a functions of redshift can be used to infer the initial conditions of the 
matter perturbations, the abundance of baryonic matter (through the baryon acoustic 
oscillations), the linear and non-linear growth of structure, as well as the mass and 
hierarchy of neutrinos (e.g., Jimenez et al., 2010). 
In this paper we present 3D power spectrum measurements of the weak lensing effect, a
statistic known as 3D cosmic shear, and use this to explore differences between the 
inferred matter power spectrum and that predicted by the standard $\Lambda$CDM model as set by the latest CMB data.
3D cosmic shear is complementary to galaxy clustering measurements of the matter power spectrum 
that can be affected by the potentially biased mapping between the galaxy distribution and 
the underlying dark matter distribution. 

There are several ways in which the weak lensing signal can be 
used to infer cosmological parameters. 
The most popular method to be applied to data is a real (configuration) space measurement of the 
2-point statistics of the data, 
a correlation function, of the galaxy ellipticities either on an assumed 2D plane or in a series of 
2D redshift slices (where inter-slice and intra-slice correlations are performed) that is referred to as 
`tomography' (Hu, 1999). This approach has a complicated scale and redshift-dependent sensitivity to the matter power 
spectrum (see e.g. MacCrann et al., 2014; Kitching, Heavens, Miller, 2011). Like all correlation function-based approaches, 
it does not offer a clear separation of linear versus non-linear scales which is more natural in Fourier-space.
Depending on the choice of weight functions used, the observations need to be tested against 
predictions that necessitate accurate modelling to very small scales down to $\sim 300$kpc or less. 
On these scales poorly known effects are dominant, making accurate cosmological parameter inference extremely challenging. 
Such configuration-space based measurements on recent data from the CFHTLenS (Erben et al., 2013; Heymans et al., 2013) 
survey have been shown to be statistically inconsistent/discrepant 
(colloquially referred to as being ``in tension'') with 
recent measurements of the matter clustering from the Cosmic Microwave Background (CMB; Planck, 2013). 
Within a standard,  
power-law $\Lambda$CDM model, the value of the variance of the linear matter perturbations on $8 h^{-1}$Mpc scales, 
$\sigma_8$, inferred from  the weak lensing correlation functions measurements is lower than that inferred from  the CMB.

There have been several studies (Battye et al. 2015, MacCrann et al. 2014, Dossett et al. 2015, Joudaki
et al., 2016) attempting to determine the cause of this discrepancy by adding additional parameters to the 
likelihood analyses which describe both systematic effects in the data or in the analysis and new physics. 
In this paper we use an alternative 3D power spectrum approach: 3D cosmic shear. 

The 3D cosmic shear method uses the 3D spherical-Bessel representation of the weak lensing galaxy ellipticities as data. 
The covariance of this data -- the 3D power spectrum -- is the quantity that contains the cosmological information.
This statistic was introduced by Heavens (2003) and developed by Castro, Heavens, Kitching (2005); 
Heavens, Kitching, Taylor (2006); Kitching (2007), Kitching, Taylor, Heavens, (2008); and 
Kitching, Heavens, Miller (2011). It was a applied to a small data set in Kitching et al. (2007), 
and then on a wide-field data set in Kitching et al. (2014) where several improvements to the method were also 
presented; including the splitting of the signal into E and B-mode components, the application of a pseudo-$C_\ell$ analysis accounting 
for the mask in the data, and the extension of the method to include the correct correlations between the real and imaginary 
parts of the theoretical covariance. An investigation of the scale-dependency of the statistic was also presented, where 
it was shown that, by making simple scale-cuts in the data vector and theory, a self-consistent set of scales can be 
defined to which the signal is sensitive over all redshifts. 
This property makes the 3D cosmic shear approach robust to effects which are strongly scale-dependent or 
localised in certain $k$ scales, such as strong non-linearities. 
In Kitching et al. (2014) the data set used was again CFHTLenS and it was found that when only using large scales in 
the statistic, more than $\sim 1$ Mpc, results were consistent with the CMB Planck data -- albeit with larger error-bars -- 
but when including smaller 
scales of $\sim 0.2$--$1$ Mpc results were no longer consistent. On small-scales it was found that the amplitude of 
matter clustering parameterised by $\sigma_8$ was lower than that measured from Planck at a significance of more than 
$2\sigma$.

Finding an explanation for this discrepancy with the Planck data is necessary, 
since if it were real it could be an important signature of new physics. 
In this paper we explore the reason for this discrepancy 
by extending the analysis and the modelling presented in Kitching et al. (2014). In particular we make several 
improvements to the statistic (as a result of computational software and hardware improvements) that allow for ten times 
more angular modes, and twice as many radial modes to be included in the analysis; this results in a higher total signal-to-noise, 
and therefore better cosmological constraints, and an increased resolution in the angular and radial directions. We also 
extend the calculation to include intrinsic galaxy alignment effects (see e.g. Joachimi et al., 2015 for a review), and 
we test the method more extensively on simulated data that includes simulated 
masks, to show that the pseudo-$C_\ell$ approach does not introduce biases in the cosmological parameters. 
We extend the 
cosmological model that is fitted to the data to include the possibility of massive neutrinos, and also include a parameterisation
for small-scale departures from the dark matter-only power spectrum caused by the presence of baryons. Finally we include 
systematic nuisance parameters to encode potential photometric redshift biases.  

In this paper we will pay particular attention to the scale-dependence of changes in the matter power spectrum 
on small-scales $k\simeq 1.5$-$5h$Mpc$^{-1}$ (physical scales of $\sim 1$Mpc). 
The power spectrum can be delimited into various regions as a function of scale that reflect the 
dominant physics at play which must be included to model its functional form: 
on the very largest scales $k<0.1h$Mpc$^{-1}$
the amplitude of matter clustering is dominated by linear physics evolving the initial primordial density fluctuations in 
the early universe; on intermediate scales $k\sim 0.5$-$1h$Mpc$^{-1}$ gravitational collapse 
of the dark matter dominates, 
this is a non-linear process but can be investigated using analytical techniques and N-body simulations; then on the smallest scales of 1 Mpc and less in the highly non-linear regime ($k>1h$Mpc$^{-1}$) non-gravitational effects driven by the
baryonic content of the Universe may begin to dominate. This effect is expected to develop as galaxy evolution progresses, 
with the peak of the star formation rate occurring at redshifts of approximately $z\simeq 2$. 
Hence, the small-scale power spectrum is very poorly understood at the current time for three reasons. 
The first is that 
dark matter clustering is not well modelled: current simulations are only precise to a few percent up to scales of 
$\sim 1$ Mpc, but not below (e.g., Lawrence et al., 2010). 
The second is that the $\Lambda$CDM paradigm could break down at small 
scales and new physical processes could be present, 
for example some modified gravity models, neutrino physics, and warm dark matter models have signatures at 
scales smaller than $1$Mpc. 
The third is that baryonic feedback processes may dominate on scales smaller than 
$1$ Mpc (e.g. van Daalen et al, 2011). 
Of these problems the baryonic feedback process is the least well understood. On scales of 1 Mpc and less, stars, 
galaxies and other baryonic components of the Universe can affect the dark matter clustering, in an unknown way. 
White (2004) provided a simple model to elucidate the effects of baryonic cooling on predictions of the power 
spectrum for weak gravitational lensing; and predicted that percent level effects may be seen. Zhan \& Knox (2004) 
provided a mixed dark matter-baryon model that included effects of baryonic cooling and the inter-cluster medium 
they also found that the weak lensing power spectrum would be impacted by a few percent. Jing et al., (2006) 
ran a set of N-body and hydro-dynamical simulations to attempt to model the impact of baryons and found 
that up to a 10\% effect could be caused on the weak lensing power spectrum. Zentner, Rudd \& Hu (2008), building 
on the N-body simulations from Rudd et al., (2008) proposed that the problem of baryonic feedback 
could be mitigated by self-calibrating weak lensing surveys i.e. adding additional (nuisance) parameters 
to model the impact of baryons. They used a simple toy model where only the concentration of 
dark matter haloes was changed, and found that cosmological parameters could be biased by up to 40\% using 
even this simple model. Mead et al. (2015) also use a physically-motivated model based on the modification to halo profiles.
A significant advance was made when Schaye et al. (2010) and van Daalen et al. (2011) used the N-body and 
hydrodynamical simulations called OWLS (OverWhelmingly Large Simulations) that also included AGN feedback. 
They found that the addition of AGN could have up to a 20\% effect on the matter power spectrum at  
$k\gs 5h$Mpc$^{-1}$, other mechanisms have smaller effects, around a few percent. 
Therefore there are at least three effects: baryonic cooling, the effects of the intra-cluster medium and AGN. However this is by no means 
an exhaustive list, for example hyper-novae may also impact the dark matter clustering, and each of these 
are not isolated effects: feedback between these effects may also be important. 
In this paper we present a flexible non-parametric approach for extracting small-scale power spectrum variation from 
N-body simulations and apply this to the Illustris (Nelson et al., 2015) and OWLS simulations. We then use 
the functions and parameters 
determined by this method, as additional degrees of freedom in the likelihood analysis of the data using 3D cosmic shear. 

This paper is structured as follows. In Section \ref{Methodology} we present the method and approach, 
in Section \ref{Results} we present results and discussion, and in Section \ref{Conclusion} we present conclusions. 

\section{Methodology}
\label{Methodology}
We refer to Kitching et al. (2014) for a exposition of the analysis in this paper, 
and also to Kitching, Heavens, Das (2014) for the inclusion of intrinsic galaxy alignment effects. 
We only restate the main points of this formalism here, and refer the reader to these papers for a full 
and more detailed presentation of the method. 

\subsection{Formalism}
We use a 3D spherical-Bessel representation of the galaxy 
ellipticity field where the transform coefficients computed on the data are 
\be 
\label{tfcoeffs}
e_{\boldsymbol{\ell}}(k)=\sum_g e_g(\boldsymbol{\theta}, r) j_{\ell}(kr) {\rm e}^{-{\rm i}\boldsymbol{\ell}.\boldsymbol{\theta}} 
\ee
where $k$ is a radial wavenumber, $\boldsymbol{\ell}$ is an angular wavenumber, $\boldsymbol{\theta}$ and $r$ are vector angular 
and radial coordinates respectively with $r$ being a comoving distance, the $j_{\ell}(kr)$ are 
spherical-Bessel functions, with $\ell=|\boldsymbol{\ell}|\gg1$. Flat sky is assumed. This is a sum over 
all galaxy ellipticities $e_g(\boldsymbol{\theta},r)$ in a data set set, 
labelled $g$, that are complex (spin-2) quantities $e_g=e_{1,g}+{\rm i}e_{2,g}$. 
The resulting four transform coefficients are complex quantities, that can be weighted by $\ell$-mode combinations 
to separate out the transform coefficients that relate to the 
$E$ and $B$-mode components of the ellipticity field $e^E_{\ell}(k)$ and $e^B_{\ell}(k)$, 
and also to remove the effects of any multiplicative 
systematic effect in the data measurements, as described in Kitching et al. (2014) Appendix A. 
  
The mean of these transform coefficients is zero, but the covariance is not and it is this that contains 
the cosmological information. The likelihood for parameters of interest $\psi$, assumed to be Gaussian, can be 
written as 
\be 
L(\phi)=\sum_{\ell}\frac{1}{\pi^2 |A_{\ell}|^{1/2}}
{\rm exp}\left[-\frac{1}{2}\sum_{ij}Z_{\ell}(i) A^{-1}_{\ell}(i,j)Z^T_{\ell}(j)\right]
\ee
the labels $i$ and $j$ run over a range $\{k_{\rm min},...,k_{\rm max}\}$ 
where $k_{\rm min}$ and $k_{\rm max}$ are the minimum and maximum $k$-mode values; so that for 
$N_k$ elements in the $k$-mode range the sums are over $2N_k$ modes.
$Z_{\ell}(i)=(e^E_{\ell}(k), e^{E*}_{\ell}(k))^T$;  is a concatenation of 
$e^E_{\ell}(k)$ and $e^{E*}_{\ell}(k)$, both of which are vectors $N_k$ in length. The 
affix-covariance matrix account for the complex, and correlated, nature of the spherical-Bessel transform of the 
ellipticity field and is equal to 
\be 
A_{\ell}(i,j)=\left( \begin{array}{cc}
\Gamma & R  \\
R^T & \Gamma^*  \end{array} \right)
\ee
which is made of four blocks of $N_k\times N_k$ matrices that are 
\ba 
\Gamma_{\ell}(k,k')&=&\mathbb{R}[C_{\ell}(k,k')]+\mathbb{I}[C_{\ell}(k,k')]\nn
R_{\ell}(k,k')&=&\mathbb{R}[C_{\ell}(k,k')]-\mathbb{I}[C_{\ell}(k,k')]
\ea
where $\Gamma$ is a covariance matrix and $R$ is a relation matrix. The matrix $C_{\ell}(k,k')$ is the 
complex covariance of the predicted signal (predicted covariance of the $E$-mode spherical-Bessel transform 
coefficients), which is a combination of signal and noise terms 
\be 
C_{\ell}(k,k')=\widetilde{C}_{\ell}(k,k')+N_{\ell}(k,k')
\ee
where the noise term $N_{\ell}(k,k')$ is given by equation (3) in Kitching et al. (2014). The signal part is a 
pseudo-$C_\ell$ estimator of the predicted covariance that accounts for the masking of the data through a 
multiplication with a 3D mixing matrix $M^{3D}_{\ell\ell'}$ via 
\be 
\label{pcl}
\widetilde{C}_{\ell}(k,k')=\left(\frac{\pi}{2}\right)^2\sum_{\ell'}\left(\frac{\ell'}{\ell}\right)
M^{3D}_{\ell\ell'} C^S_{\ell}\left(k\frac{\ell'}{\ell},k'\frac{\ell'}{\ell}\right).
\ee
The original signal covariance $C^S_{\ell}$ can be derived using the relationship between the 
lens potential and the Newtonian potential integrated along the line of sight, and linking the 
Newtonian potential to the underlying matter perturbations via Poisson's equation. 
The dependence on cosmological parameters comes through the $C^S_{\ell}$. This results in a 
predicted complex covariance that is a combination of terms from the intrinsic galaxy ellipticity and additional 
cosmic shear. 
 
The observed ellipticity is a combination of the intrinsic (unlensed) galaxy ellipticity $e^I$ and the additional 
ellipticity caused by the weak gravitational lensing along the line of sight called shear $\gamma$. In the case that 
$|\gamma| \ll |e^I|$ then the observed ellipticity is a linear sum of these quantities $e=e^I+\gamma$, which 
means that when taking the covariance of the observed shear transform coefficients the result is four terms that 
correspond the quadratic combination of the intrinsic ellipticity and the shear (see Kitching, Heavens, Das, 2014) 
\ba 
C^S_{\ell}(k,k')&=&C^{\gamma\gamma}_{\ell}(k,k')+C^{II}_{\ell}(k,k')\nn
&+&C^{I\gamma}_{\ell}(k,k')+C^{\gamma I}_{\ell}(k,k').
\ea
Here the superscript refers to the terms that are included in each covariance. The last term - the 
correlation between a foreground galaxy's observed shear and a background galaxy's intrinsic ellipticity - is 
expected to be zero by construction, but we include it in all calculations as redshift uncertainty can reverse the order of the assumed distances, and cause the 
observed correlation to be non-zero. The power spectrum for quantities $X$ and $Y$, 
which in this case are either $I$ or $\gamma$, can be written as a matrix multiplication 
\be 
C^{XY}_{\ell}=\widetilde{G}^X_{\ell}\widetilde{G}^{\dagger,Y}_{\ell}
\ee
where $\dagger$ refers to a transpose and complex conjugate and the matrices $\widetilde{G}^X_{\ell}$ are 
\be 
\widetilde{G}^X_{\ell}(k,k')=D {\mathcal A}\frac{(\Delta k)^{1/2}}{k} G^X_{\ell}(k,k')
\ee
where $\Delta k$ is a resolution in the radial wavenumber that approximates an integral, $D=D_1+{\rm i}D_2$ 
is a complex variable where $D_1=\frac{1}{2}(\ell_y^2-\ell_x^2)$ and $D_2=-\ell_x\ell_y$, where $\ell_x$ and 
$\ell_y$ refer to the wavenumber components in the $x$ and $y$ Cartesian coordinate frame.  
${\mathcal A}=3\Omega_{\rm M} H_0^2/(\pi c^2)$ where $H_0$ is the current value 
of the Hubble parameter, $\Omega_{\rm M}$ is the ratio of the total matter density to the 
critical density, and $c$ is the speed of light in a vacuum. The $G$ matrices are different for the 
intrinsic and shear parts of the covariance. 

For shear the $G$ matrix is 
\be
\label{Ggamma} 
G^{\gamma}_{\ell}(k,k')=\int {\rm d}z_p{\rm d}z' j_{\ell}(kr[z_p])n(z_p)p(z'|z_p)U^{\gamma}_{\ell}(r[z'],k')
\ee
where $n(z_p){\rm d}z_p$ is the number of galaxies in
a spherical shell of radius $z_p$ and thickness ${\rm d}z_p$, $p(z'|z_p)$ is
the probability of a galaxy at redshift $z'$ to have a photometric redshift 
$z_p$, $j_{\ell}(kr)$ are spherical Bessel functions. 

The matrix $U$ is 
\be 
\label{Ueq}
U^{\gamma}_{\ell}(r[z],k)=\int_0^{r[z]}{\rm d}r'\frac{F_K(r,r')}{a(r')}j_{\ell}(kr')P^{1/2}(k;r')
\ee
where $P(k; r)$ is the matter power spectrum at comoving
distance $r$ at radial wavenumber $k$; we refer the reader
to Castro, Heavens, Kitching (2005) for a discussion of
the approximation involved in using the square-root of the
power spectrum here. $F_K = S_K(r - r')/S_K(r)/S_K(r')$ is the lensing kernel
where $S_K(r) = \sinh(r)$, $r$, $\sin(r)$ for cosmologies with spatial curvature 
$K = -1$, $0$, $1$, and $a(r)$ is the dimensionless scale factor at the cosmic 
time related to the look-back time at comoving distance $r$.
The combination of the $G$ and $U$ matrices create the
covariance of the $\gamma_E(k,\ell)$ spherical-Bessel transform coefficients where
$C^{\gamma\gamma}_{\ell}(k, k')=\langle {\mathbb R}[\gamma_E(k, \ell)]{\mathbb R}[\gamma_E(k', \ell)]\rangle$; 
the same expression is true for
imaginary parts ${\mathbb I}[\gamma_E(k,\ell)]$ and in 
the likelihood both terms are contributors. 
Throughout this investigation we 
use {\sc camb}\footnote{{\tt http://camb.info} version 2012.} to calculate 
the matter power spectra with the {\sc halofit} (Smith et al., 2003) non-linear
correction and the module for Parameterized Post-Friedmann (PPF) prescription for the dark energy
perturbations (Hu \& Sawicki, 2007; Fang et al., 2008; Fang, Hu \& Lewis, 2008)\footnote{
 This is to be consistent with the Kitching et al., (2014) analysis, although we 
do not actually vary the dark energy equation of state in this paper.}.

For the unlensed part of the galaxy ellipticity, we use the linear alignment model of 
Hirata \& Seljak (2004), where the intrinsic galaxy ellipticity 
is linearly related to the local second derivative of the primordial Newtonian potential. 
This propagates through to a spherical-Bessel covariance, as 
described in Kitching, Heavens, Das (2014). In this case the $G$ matrix is 
\be 
G^I_{\ell}(k,k')=\int {\rm d}z_p{\rm d}z' j_{\ell}(kr[z_p])n(z_p)p(z'|z_p)\frac{U^{I}_{\ell}(r[z'],k')}{r^2[z']}
\ee
where
\be 
\label{eIA1}
U^I_{\ell}(r[z],k)=\int_0^{r[z]}{\rm d}r'\frac{\delta^D(r'-r)I(r'[z])}{a(r')}j_{\ell}(kr')P^{1/2}(k;r')
\ee
and the factor $I(z[r])$ is 
\be 
\label{eIA2}
I[r(z)]=\left(\frac{c^2}{2 H^2_0}\right)\left(\frac{2.1\times 10^{-3} A_{\rm IA}}{\bar D(z)}\right).
\ee
$A_{\rm IA}$ parameterises the amplitude of the intrinsic alignment signal, which has been used in several 
forecasting papers (e.g. Kirk et al., 2015), and also fit to data using correlation function 2-point statistics (e.g. 
Heymans et al., 2013). The 
$U$ matrices for both the shear and intrinsic signal effectively encapsulate the redshift kernel of the signal, 
where the lensing geometric kernel can be seen in the shear case - the effect being a distance-weighted integral 
along the line of sight, and a localised delta-function in the intrinsic alignment case. $\bar D(z)$ is the 
linear growth factor. 

\subsection{Implementation}
The above formalism is coded in a software {\tt 3dfast}\footnote{The code is available here 
{\tt https://github.com/tdk111/3dfast}.}, which was used in Kitching et al. (2014).  
In this paper we present an improved analysis, as a result of software and hardware 
improvements used for the cosmological parameter inference. The main result of this is an increase 
in the number of $\ell$ and $k$-modes available for the analysis. In Kitching et al. (2014) 
$164$ independent angular modes were used. In this paper this is increased by a factor of 
$10$ to $1640$ independent $\ell$-modes over the range $\ell_{\rm min}=360$ to $\ell_{\rm max}=4970$. 
In the radial direction we use $50$ $k$-modes linearly sampled between 
$0.001-5h$Mpc$^{-1}$, for each $\ell$-mode.  We choose 
$k_{\rm max}=5h$Mpc$^{-1}$ to avoid the extremely non-linear regime of less than 
a few hundred kiloparsecs in comoving seperation (see Section \ref{Introduction}). 
This leads to $82$,$000$ modes measured from the data, and 
$4.1\times 10^6$ modes to be modelled in the covariance\footnote{The current implementation of {\tt 3dfast} can 
compute one covariance matrix for this dataset in $\sim 10$ seconds on node $36$ of this machine 
{\tt http://hipatia.ecm.ub.es/ganglia/}.}. This choice of angular modes avoids large scales, of more than one degree. 
For the spherical-Bessel shear transform our $\ell$-mode selection corresponds an 
angular range of $4$-$60$ arcminutes;  and this mapping from $\ell$-mode to real-space angle 
is unaffected by the choice of $k$-modes due to the orthogonality properties of the spherical-Bessel transform.
On scales larger than this Asgari et al. (2016) use a correlation 
function approach (COSEBIs), and map a $k$-mode and redshift-dependent angular range onto $\ell$, finding that 
$\ell=360-5000$ in that analysis corresponds to $40-100$ arcminutes, and in doing so find a 
signature of $B$-modes in the CFHTLenS data over those configuration-space angular scales. We avoid such scales in this 
analysis, and note that a full comparison between COSEBIs and spherical-Bessel weighting requires further investigation.

\begin{figure*}
    \includegraphics[angle=0,clip=,width=0.66\columnwidth]{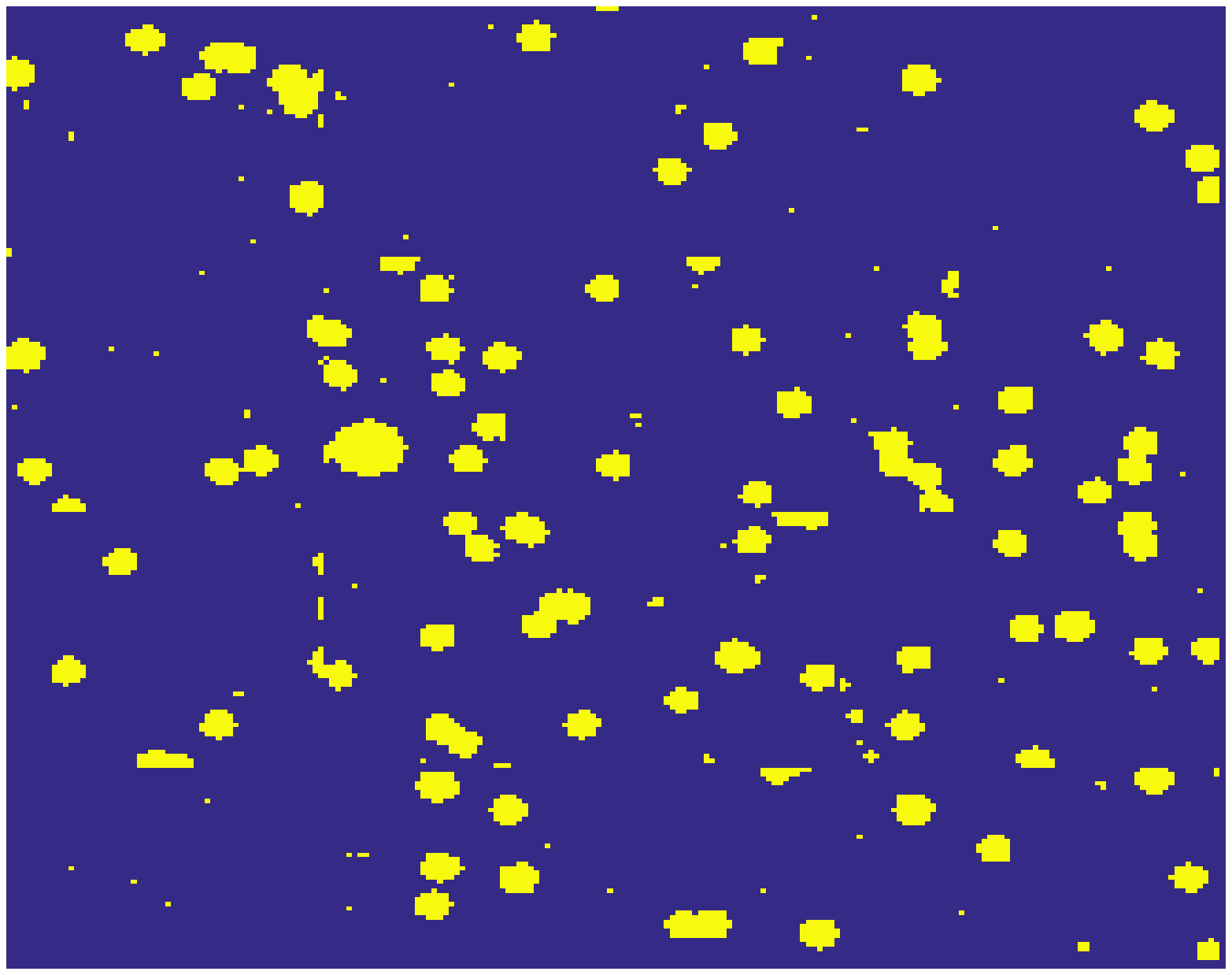}
    \includegraphics[angle=0,clip=,width=0.66\columnwidth]{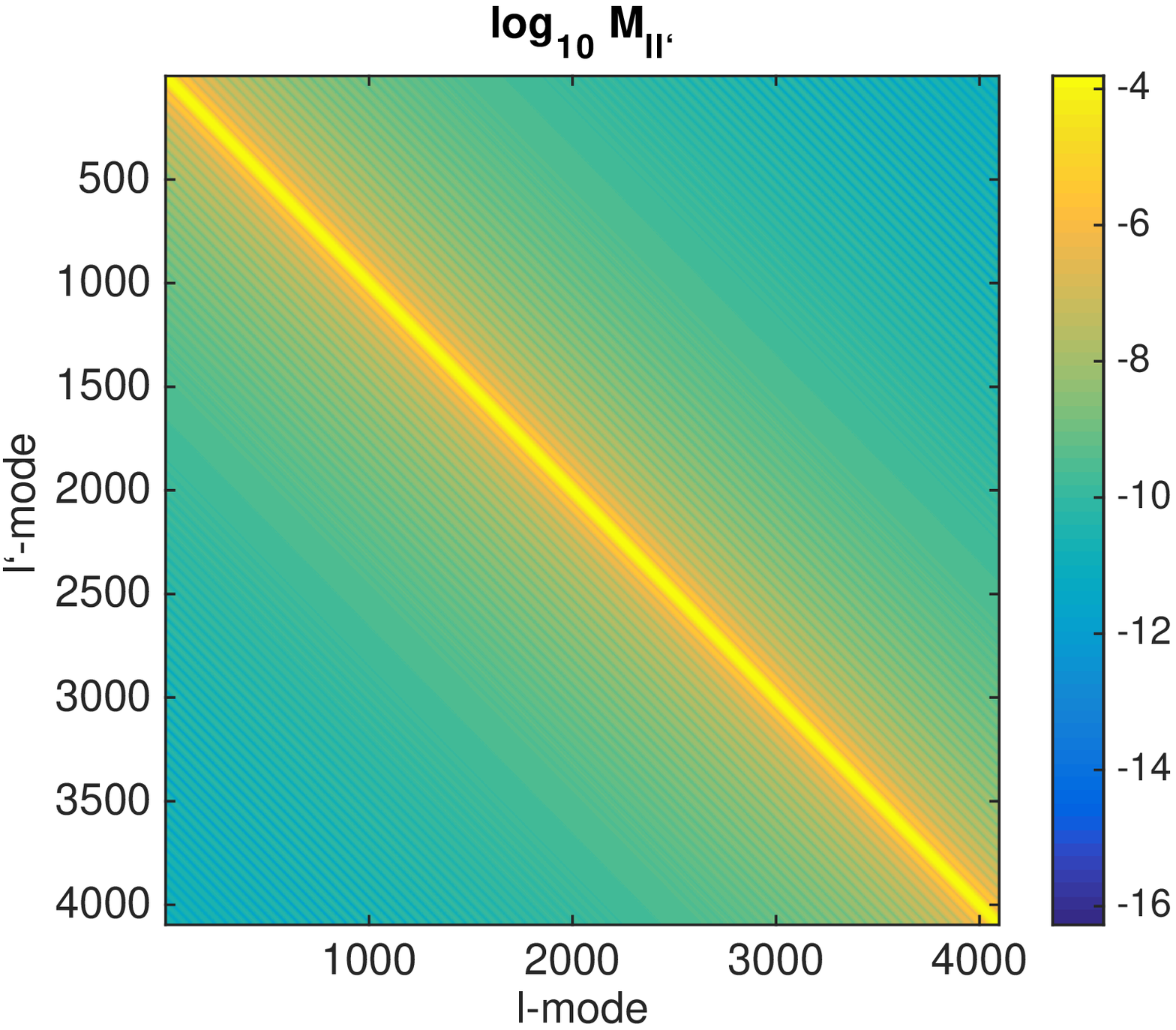}
    \includegraphics[angle=0,clip=,width=0.66\columnwidth]{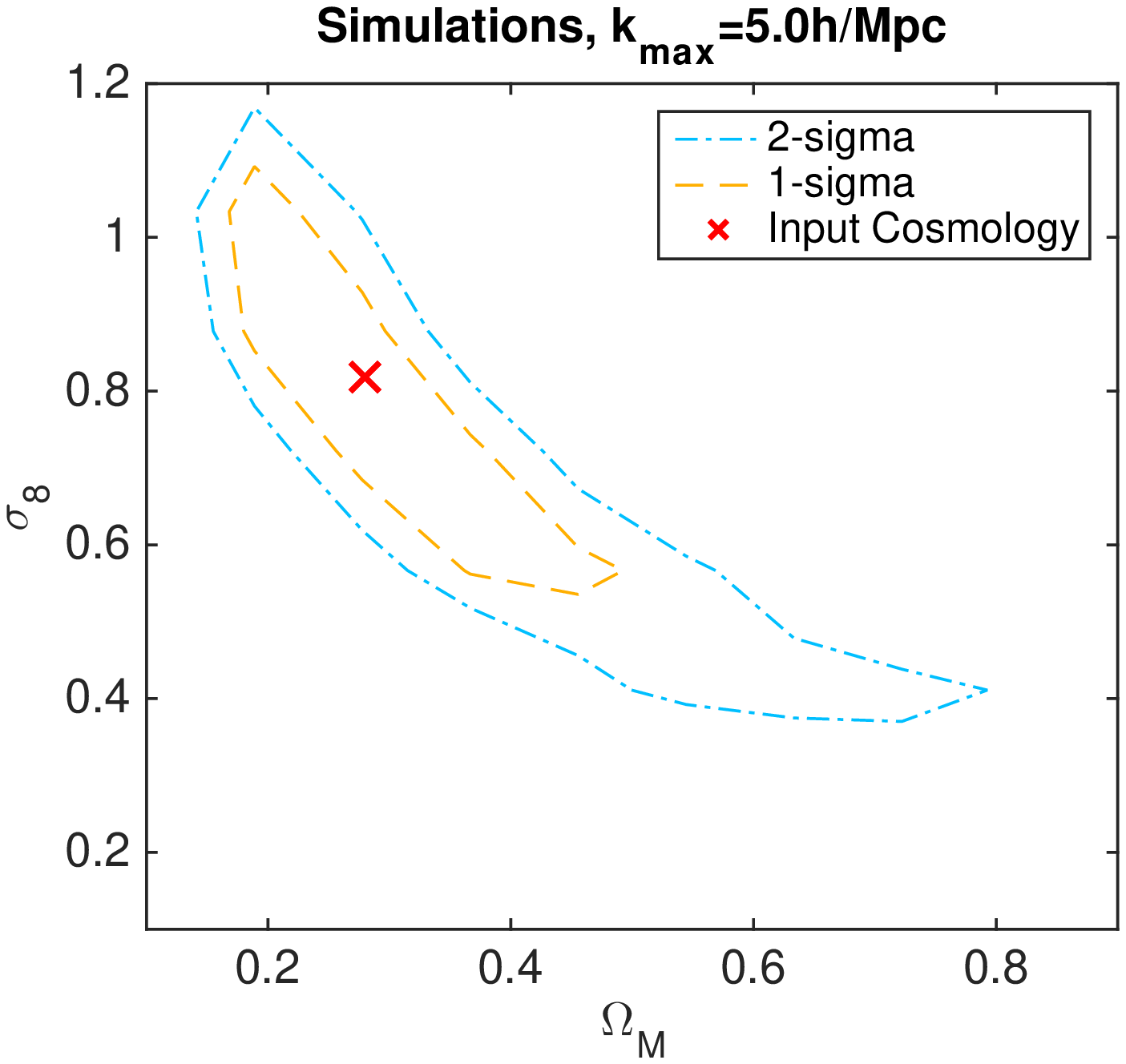}
 \caption{Lefthand panel: A typical mask in the CLONE simulations (from the line-of-sight $1$, showing $12.84$ 
square degrees), where yellow are the masked 
regions that simulate stellar masks.
Middle panel: A typical derived mixing matrix from the simulations, where we use a maximum $\ell$-mode of $4000$.
Righthand panel: The mean $1\sigma$ (dashed) and $2\sigma$ (dot-dashed) contours averaged over all the simulations 
in the $(\sigma_8$, $\Omega_{\rm M})$ plane, compared to the input cosmology (the red cross).}
\label{clone}
\end{figure*}

To test the implementation we use the CFHT N-body CLONE simulations (Harnois-D\'eraps et al., 2012). 
These simulations were made assuming a flat $\Lambda$CDM cosmology with parameters 
$\Omega_{\rm M}=0.279$, $\Omega_{\rm B}=0.046$, $n_s=0.96$, $h=0.701$ and $\sigma_8=0.817$. 
Whilst not being fully 3D simulations, 
they are finely binned in redshift with $26$ bins over the range $z=0-3$. In each redshift bin the matter density is 
projected onto a 2D plane. There are $184$ independent lines of sight, where in each one weak lensing shear information is generated 
via ray tracing through the simulations. Importantly 
these simulations have realistic masking, and are tailored to mimic the survey number density, geometry, 
and noise properties of the CFHTLenS survey; which is the data set we use in this paper. The presence of 
the realistic masks means that the pseudo-$C_\ell$ mask-correction can be tested. In addition we supplement the CLONE simulations 
with realistic photometric redshift posterior probabilities: we take the photometric redshifts posterior probabilities 
from CFHTLenS, and then assign a posterior to each CLONE galaxy with the appropriate mean redshift, the best estimated photometric 
redshift is then re-sampled from the assigned posterior. 
In Figure \ref{clone} we show the result of applying the current implementation to the simulations where 
we split the available lines of sight, each of $12.84$ square degrees, into groups of $12$ that are approximately the same 
total area as the CFHTLenS survey, which is $154$ square degrees (this leaves a remainder of $4$ simulations, lines-of-sight 
$180-184$, which we do not use), to create simulated data of the same size 
as used in this paper. We show the 2-parameter likelihood contours in the 
$(\sigma_8$, $\Omega_{\rm M})$ plane, marginalised over $\Omega_{\rm B}$, $h$ and $n_s$ in a flat $\Lambda$CDM cosmology 
(see Section \ref{Model Parameters}). We find that the likelihood analysis recovers the input cosmology in all cases. 

\subsection{Data}
The data we use is the CFHTLenS data (Erben et al., 2013; Heymans et al., 2013), which is a $154$ 
square degree optical survey (over four fields W1, W2, W3, W4) 
in $ugriz$ bands, with weak lensing shape measurement (Miller et al., 2013) and 
photometric redshift posterior probabilities (Hildebrandt et al., 2012). We use the publicly available catalogues, 
and remove those fields that have been assessed to be unsuitable for cosmic shear analysis (Heymans et al., 2013) 
using star-galaxy cross-correlation statistics. This is the same data set that was used in Kitching et al. (2014). 

We follow Kitching et al. (2014) in selecting only photometrically 
identified early-type galaxies for our analysis that are expected to have 
small intrinsic alignment contamination. For example, Mandelbaum et al. (2011) found a 
null intrinsic alignment signal in the WiggleZ data 
whose selection function resulted in a galaxy sample that is similar to that of CFHTLenS. In addition 
the linear alignment model that we use is only expected to be appropriate for early-type galaxies 
(see e.g. Joachimi et al., 2015).

\begin{figure*}
    \includegraphics[angle=0,clip=,width=2\columnwidth]{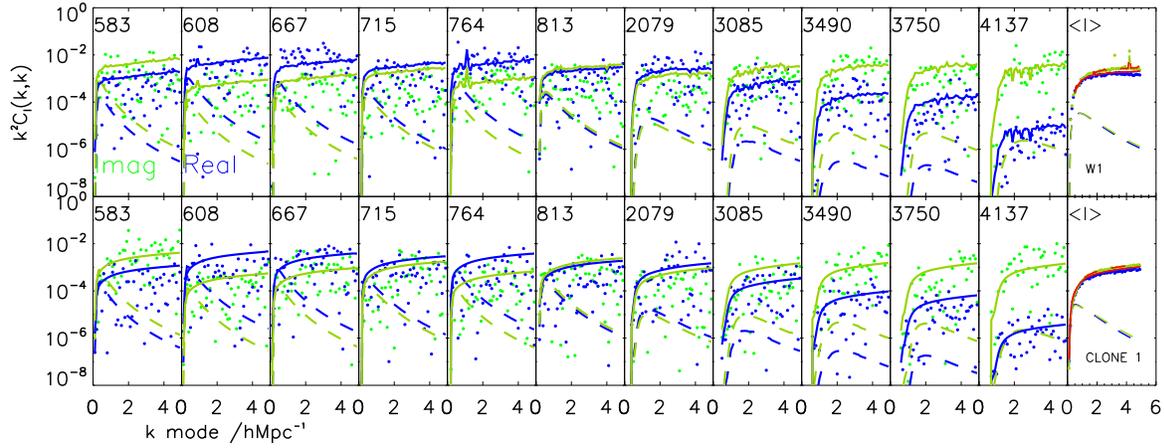}
 \caption{The real and imaginary parts of the transform coefficient data vector (blue and green points respectively) as 
a function of 
$k$-mode for a selection of the $1640$ $\ell$-modes used in the analysis, 
compared to the diagonal part of the expected pseudo-$C_\ell$ 
covariance matrix which is a sum of the noise part (solid lines) and the signal 
part (dashed lines). The rightmost panels show the mean over all $\ell$-modes used in the analysis. 
The top panels show this for the 
CFHTLenS W1 field, the bottom panels show this for the CLONE simulations line-of-sight $1$.}
 \label{transform}
\end{figure*}

In Figure \ref{transform} we show the real and imaginary measured transform coefficients (equation \ref{tfcoeffs}) 
for a selection of $\ell$-modes as a function of $k$-mode, and also the predicted diagonal of the pseudo-$C_\ell$ 
covariance $\widetilde{C}_{\ell}(k,k)$ (equation \ref{pcl}), for the CFHTLenS W1 field. We also show the same plot 
for one of the simulations used to test the pipeline. 

Because the data we use is a 1-point estimator, and the covariance that contains the cosmological information 
is analytic, there is no need to estimate the covariance matrix from 
simulations (see e.g Taylor, Joachimi, Kitching, 2014). The primary assumption in the likelihood analysis is that the
likelihood function is Gaussian, i.e. that the shear transform coefficients are Gaussian distributed. As shown in Kitching 
et al. (2014) this is a good approximation for the CFHTLenS data. This is also expected from the central limit theorem 
because if each galaxy has a posterior probability for the observed ellipticity $p_g(e)\ast p(\gamma)$ (where $\ast$ is a 
convolution) then the probability distribution of the shear transform coefficients, via equation (\ref{tfcoeffs}), 
will be $p(e_{\ell}[k])=\bigotimes_g [p_g(e)\ast p(\gamma)] j_{\ell}(kr){\rm e}^{-{\rm i}\ell.\theta}$ i.e. 
where $\bigotimes_g$ is a 
series convolution over all galaxies weighted by the spherical Bessel function, which through 
the central limit is expected to result in a Gaussian distribution.  

\subsection{Model Parameters}
\label{Model Parameters}
The model parameters that we fit to the data consist of three parts that capture the cosmological model, the 
baryonic feedback model, and the parameters for photometric redshift systematic effects. 
We adopt  as the  baseline, the set of  cosmological parameters of the flat $\Lambda$CDM model: 
$\Omega_{\rm M}$, $\Omega_{\rm B}$ and $\Omega_{\rm DE}$, the dimensionless 
densities of matter, baryons and dark energy respectively, where we always assume a flat geometry i.e. that 
$\Omega_{\rm DE}=1-\Omega_{\rm M}$; the dark energy equation of state parameter $w$, that we assume to be 
constant with redshift; the spectral index of of the initial density perturbations $n_s$; the dimensionless 
Hubble parameter $h=H_0/(100$kms$^{-1}$Mpc$^{-1})$; the variance of matter perturbations on $8h^{-1}$Mpc scales, 
$\sigma_8$; and the total sum of neutrino masses, $m_{\nu}$ for which we assume an inverted hierarchy throughout 
(the results are not sensitive to the choice of hierarchy for a data set of this size). 
In our investigations we will use the Planck Collaboration (2013) best fit parameters to fix any cosmological parameters that we 
do not explicitly vary in the analysis, and all other parameters not listed here are also fixed at these values.  
Beyond the cosmological parameters we consider ``systematic'' parameters (variables that parameterise 
systematic effects).
We include the intrinsic alignment parameter $A_{\rm IA}$ as a free parameter, where we use the 
non-linear power spectrum in the linear alignment model; this is an \emph{ad hoc} modelling of small-scale 
intrinsic alignment behaviour (see Joachimi et al., 2015; Blazek et al., 2015) but is a good empirical fit to 
galaxy-galaxy lensing data.
For other systematic parameters we focus only on those that are most likely to have an impact on small-scales. 
These are the impact of photometric redshifts, because photometric redshifts damp 
power and correlated $k$-modes on small radial scales less than the redshift error 
i.e. $k\gs 2\pi/[3000\sigma_z(z)]$, where $3000\sigma_z(z)$ is 
approximately the comoving distance error caused by photometric redshift uncertainties at a 
redshift of unity (see Kitching, Heavens, Miller, 2011 
where this is explored in more detail); and 
baryonic feedback processes that can impact scales of $k\gs 1h$Mpc$^{-1}$ (see Section 1). 

\subsubsection{Photometric Redshift Systematics}
As shown in Choi et al. (2015) there is evidence from galaxy-weak lensing cross correlations that the photometric 
redshifts in CFHTLenS are biased with respect to their (true) spectroscopic redshifts. We find that 
their bias as a function of spectroscopic redshift is well-parameterised by a linear relation, we estimate this relation 
from their tabulated results to be $z_{\rm bias}(z_s)=p_2 (z-p_1)$ where $p_2=-0.19\pm 0.05$ and $p_1=0.45\pm 0.05$. 
To model the effect of possible redshift biases we include this redshift bias function in our analysis by 
shifting the CFHTLenS photometric redshift posterior distributions by this factor in equation (\ref{Ggamma}), 
$p(z'|z_p)\rightarrow p(z'-z_{\rm bias}|z_p)$ and letting $p_1$ and $p_2$ be free parameters; which to first order is a shift 
in the mean of the function. 

With more data, a more complex bias function could be explored, but the limited statistical power of 
this dataset does not warrant this. As shown in Kitching et al. (2014) the CFHTLenS 
data set is not large enough to support parameter estimation over more than $\sim 4-5$ well constrained free parameters.  

\subsubsection{Baryonic Feedback Models}
We start by  using the results from the  OWLS OverWhelmingly Large Simulations (van Daalen et al., 2011; 
Shaye et al., 2010), 
a suite of large, cosmological, hydrodynamical simulations, which include 
various baryonic processes including AGN feedback, supernovae feedback, cooling etc. 
Their code uses a TreePM algorithm to efficiently calculate the gravitational forces and 
Smoothed Particle Hydrodynamics (SPH) to follow and evolve the gas particles.
The authors provide the matter power spectrum as a function of wavenumber $k$ and redshift $z$,  $P(k,z)$ (linking to 
equation \ref{Ueq} here we use $P(k,z)$ as a shorthand for $P(k; r[z])$ where $r(z)$ is the comoving distance at redshift $z$), 
for the same cosmology but with $9$ different baryonic effects or ``recipes''; 
their description can be found in Table 1 of van Daalen et al., (2011) (note that entries 2,5,6 are relative 
to a different cosmology and so will not be considered here).  
The large volume of the simulations means that the lowest $k$ mode sampled is 
$0.1h$Mpc$^{-1}$, reaching the (quasi)linear regime where baryonic effects are fully negligible.

In Zenter et al. (2013) the authors quantified the impact of baryonic effects 
on the  convergence power spectrum using principal component analysis (PCA; see e.g. Jolliffe 1986) 
and found that the first $2$  eigenmodes account for over 90\% of the variance among the spectra. 
Here we aim at using the same approach but for the matter power spectrum itself as a function of $k$ and $z$.

To minimise the dependence on cosmology, we choose to  model the  relative change 
induced by the baryonic effects compared to a dark matter only (DM ONLY) recipe, therefore we work with the 
quantity $R=P_i(k,z)/P_{\rm DM\, ONLY}(k,z)$ where $i$ stands for the various  baryonic recipes. We also only 
consider the  redshift range  relevant for the  present analysis i.e. $0.125 \le z \le 1.5$.
\begin{figure}
    \includegraphics[angle=0,clip=,width=0.9\columnwidth]{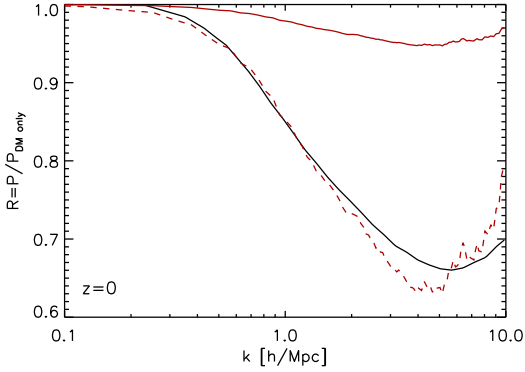}
 \caption{Comparison of OWLS and Illustris simulations 
   predictions for the  baryonic effects on the matter power spectrum $R=P(k)/P_{\rm DM \, ONLY}(k)$ at $z=0$. OWLS mean correction (i.e., $R_{\rm mean}$ in equation 15): upper red solid line; Illustris: lower black solid line; OWLS mean correction scaled by a factor 7: dashed line.}
 \label{OWLIllu}
\end{figure}

The PCA approach describes $R$ in terms of eigenvectors and eigenvalues:
\begin{equation}
R= R_{\rm mean}(k,z)+ \sum_{i=1,9}V_i(k,z) {\cal E}_i
\label{eq:PCA}
\end{equation}
where $R_{\rm mean}$ is the mean correction (the PCA-inferred mean effect of all the models considered) 
$V_i$ are the eigenvectors and ${\cal E}_i$ are the eigenvalues. We find that the second term in the RHS of equation (\ref{eq:PCA}) is of the same order of  $R_{\rm mean}$, it cannot be neglected, but cannot change 
$R_{\rm mean}$ by a large factor.
Our philosophy  then follows Eifler et al. (2015). We describe the matter power spectrum as:
\begin{equation}
P(k,z)=P_{\rm DM\, ONLY}(k,z){\cal R}_{\vec E}(k,z)
\end{equation}
with
\begin{equation}
{\cal R}_{\vec{\alpha}}(k,z)= R_{\rm mean}(k,z) + \sum_{i=1,N}V_i(k,z) \alpha_i
\end{equation}
where $R_{\rm mean}$ and  $V_i$ are  provided by the PCA procedure (equation \ref{eq:PCA} above) and  $E_i$ their coefficients 
(to be determined by the data analysis). The PCA provides $N=9$ eigenvectors but, as we shall see below, 
the first one or two already encode all the information one is interested in in this context. 
We then aim at marginalising over the coefficients of the dominant eigenmodes.
In doing so we make  2 fundamental assumptions here:  {\it 1)} that the set of 9 recipes encompasses 
all reasonable functional shape of the corrections (but not necessarily the  amplitude) and 
therefore that the set of eigenmodes that the PCA analysis will yield will be a  
full basis set for the baryonic effects (not just a full basis  set for the OWLS simulations); 
{\it 2)} that on scales larger than the largest scales modelled by the simulations the baryonic 
effects are negligible and therefore the relative effect is $0$. We will completely relax the first assumption below.

We find that  using only one PCA coefficient keeps the residuals below 0.5\% for $k < 0.5h$Mpc$^{-1}$ and 
below 1.2\% for $k < 1h$Mpc$^{-1}$; using the first two PCAs keeps the residuals below 0.1\% for $k < 0.5h$Mpc$^{-1}$.
Using no PCA coefficients, only the mean correction, we find residuals below 0.8\% for $k<0.5h$Mpc$^{-1}$ and below 
2.5\% for $k<1$. Since $R_{\rm mean}$ and $V_1$ are of about the same magnitude this means that the differences among the models are at least as big as the effect itself. The recipe therefore would be to set the mean correction and the first 
PCA eigenvector, leaving its amplitude a free parameter. One would expect the recovered parameter 
value not to be much larger than unity for the modelling adopted to be valid.

To relax our first assumption above,
we next test if this PCA description of the baryonic effects on the matter $P(k)$ shape can describe the effects found by 
an independent set of simulations. We use the Illustris simulation (Nelson et al., 2015) which incorporates a broad range of 
astrophysical processes that are believed to be relevant to galaxy formation (gas cooling, energy feedback from black holes, 
supernovae, AGN).  While gravitational forces are calculated using a Tree-PM scheme as in OWLS, the hydrodynamics 
are modelled by the moving-mesh technique (see Nelson et al., 2013).  In particular we refer to Figure 5 of 
Vogelsberger et al. (2014). We find that the  relative effect of baryons on the matter power spectrum, $R$, 
at $z=0$ is 7 to 8 times larger than it is in the mean of the OWLS effects at the same redshift. 
While OWLS had 9 baryonic recipes, in our PCA-based representation
they  are described by few $\sim$few \% eigenmodes around a ``mean'' correction of $ \sim 3$\%  at low redshifts 
at $k\sim 1h$Mpc$^{-1}$ (up to 8\% at  higher $k$). Illustris on the other hand presents only one model 
at $z=0$  with $|R-1| \sim 20$\% at $k<1 h$Mpc$^{-1}$(up to more than 35\% at larger $k$). No reasonable values of the 
OWLS-extracted PCA coefficients could reproduce such an effect; 
even the AGN model in van Daalen et al., (2011) gives only  a $\sim 10\%$ suppression at 
$k\sim 2 h$Mpc$^{-1}$. We therefore (also) explore 
a model that  can interpolate between the two simulations by 
adding  a free parameter that rescales the mean correction for OWLS. This is illustrated in 
Figure \ref{OWLIllu} where the power spectra ratio $R$ at $z=0$ are shown for the Illustris simulation, 
the mean correction from OWLS,  $R_{\rm mean}$ in eq. 15), and this correction rescaled by a factor 7. The resulting form of the function that we
fit to the data is  
\be 
R=1+[R_{\rm mean}(k,z)-1]E_1+V_1(k,z)E_2
\ee
where $E_1$ and $E_2$ are free parameters, and the resulting range of the variation in the function can capture both the 
OWLS and Illustris behaviour\footnote{The PCA data and code to read in and manipulate the functions is available 
here {\tt https://github.com/tdk111/baryonmodel}.}. Schneider \& Teyssier (2015) also present 
an investigation of baryonic feedback behaviour, whose power spectrum suppression again requires of order two components: 
a supression amplitude, and k-range at which that suppression begins to affect the power spectrum; however we have not 
tested our ability to recover their results.

\section{Results}
\label{Results}

\begin{figure*}
    \includegraphics[angle=0,clip=,width=2\columnwidth]{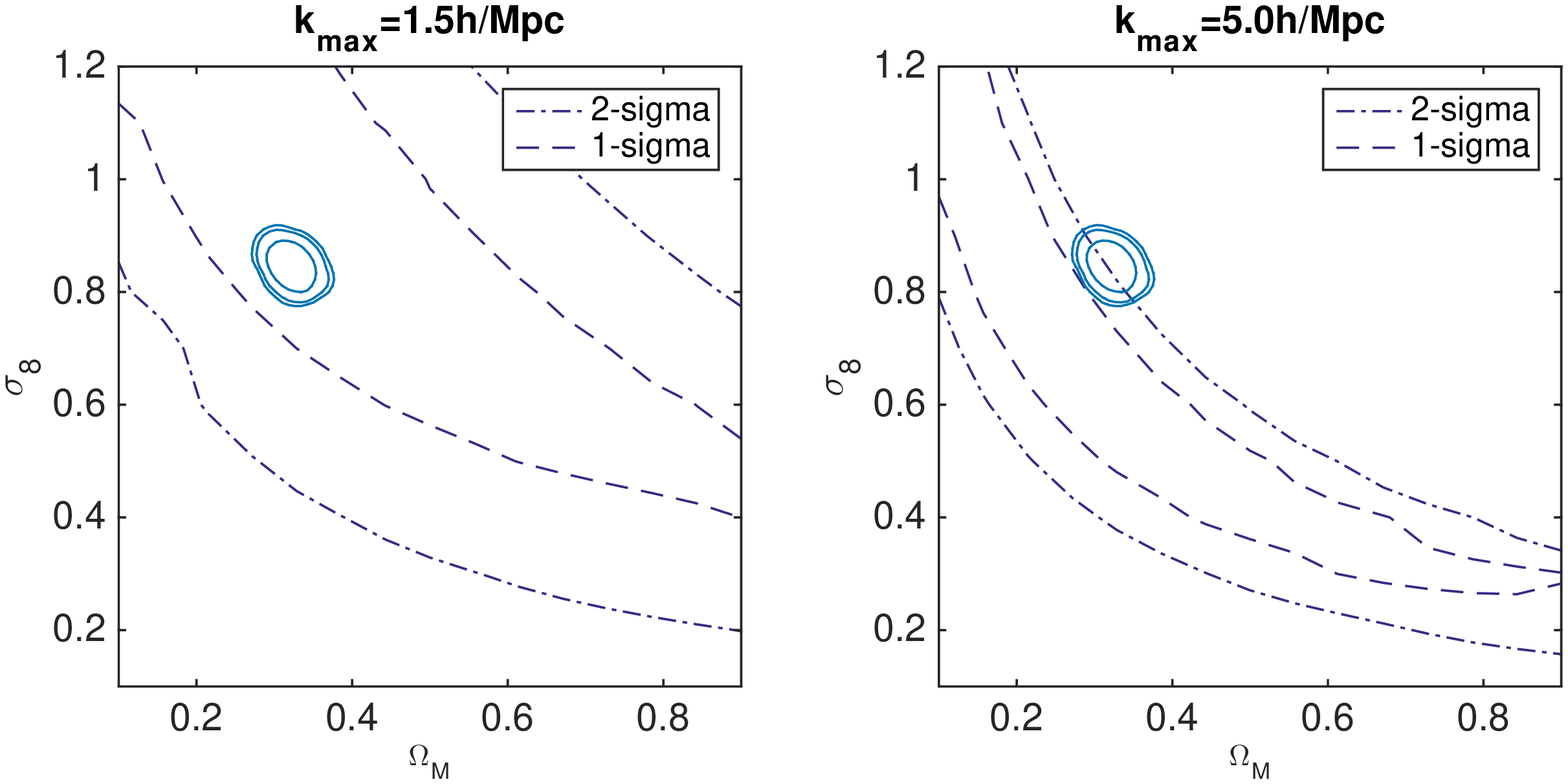}
 \caption{Projected parameter constraints for a $\Lambda$CDM cosmology in the $(\sigma_8,$ $\Omega_{\rm M})$ plane showing $1\sigma$ (dashed) and $2\sigma$ (dot-dashed) contours for maximum $k$-modes of $1.5h$Mpc$^{-1}$ and $5h$Mpc$^{-1}$ in the analysis. Note that other parameters are fixed at Planck best-fit values. The 
Planck $1\sigma$, $2\sigma$ and $3\sigma$ parameter contours are shown with the blue solid lines. 
Here baryonic effects, photometric redshift biases, intrinsic alignments and neutrino mass are all 
assumed to be zero.}
\label{datac}
\end{figure*}

We vary the free parameters in our analysis, and estimate their posterior probability distributions using a 
Metropolis-Hastings MCMC chain with a proposal distribution that is determined using the Fisher matrix of the parameters 
involved (the Fisher matrix is defined in Kitching, Heavens, Miller et al., 2011). 
We do not assume any priors on our parameters in the analysis, except very wide boundaries to prevent the MCMC chains 
from moving into unphysical parameter areas, these are $\Omega_{\rm M}>0$, $\sigma_8>0$, $h>0$, $|A_{\rm IA}|<100$.  

For illustration of the tension, in Figure \ref{datac} we show the projected $1\sigma$ and $2\sigma$ contours in the $(\sigma_8,$ $\Omega_{\rm M})$ plane using  
maximum $k$-modes of $1.5h$Mpc$^{-1}$ and $5h$Mpc$^{-1}$ in the analysis.  Note that in this figure all other cosmological parameters are fixed at the base $\Lambda$CDM Planck best fit values and the 
systematic parameters are at their fiducial values (no intrinsic alignment, no baryonic effects, no photo-z bias). 
This is compared to the 
Planck constraints\footnote{We use the Planck Legacy Archive chain {\tt PLA/base/planck\_lowl/base\_planck\_lowl\_1.txt}.} 
in the same 
plane. It can be seen that for (quasi-) linear scales the data is fully consistent with the Planck data. However there is a 
tension at small-scales. The constraints are slightly broader than those expected from the simulated data (Figure \ref{clone}), 
this is because the power spectrum signal-to-noise is lower than expected due to the lower $\sigma_8$ value. 

To investigate what could be causing the tension with the Planck constraints in the $(\sigma_8,$ $\Omega_{\rm M})$ plane we 
fixed the $\Lambda$CDM parameters at the Planck maximum likelihood values, and then only varied the 
additional parameters in our analysis to 
gauge if non-canonical values of them can explain this tension; thereby placing Planck $\Lambda$CDM-conditional constraints on these 
parameters. The additional parameters are the intrinsic alignment amplitude $A_{\rm IA}$, the sum of the neutrino masses $m_{\nu}$, 
the two baryonic feedback parameters $E_1$ and $E_2$, and the two photometric redshift bias parameters $p_1$ and $p_2$.
These parameters are all varied simultaneously in the fitting, except where we explicitly fix 
the intrinsic alignment amplitude to be zero. By fixing all other $\Lambda$CDM parameters, 
including $\sigma_8$ and $\Omega_{\rm M}$, we infer the values of the additional parameters conditional on the Planck cosmology being correct.
\begin{table*}
\begin{tabular}{|l|l|l@{ $\pm$ }l|l|}
Intrinsic Alignment& Parameter& Mean&$1\sigma$ &\\
\hline
Free & $A_{\rm IA}$& $-11.3$&$5.9$&\\
Free & $m_{\nu}$/eV& $0.13$&$0.15$&\\
Free & $E_1$& $-0.06$&$0.09$& baryon model\\ 
Free & $E_2$& $ 0.00$&$0.02$& baryon model\\
Free & $p_1$& $ 0.27$&$0.06$& photometric bias\\
Free & $p_2$& $-0.29$&$0.07$& photometric bias\\
\hline
Zero & $m_{\nu}$/eV& $0.14$&$0.12$&\\
Zero & $E_1$& $ 0.06$&$0.15$& baryon model\\
Zero & $E_2$& $ 0.00$&$0.02$& baryon model\\
Zero & $p_1$& $ 0.26$&$0.05$& photometric bias\\
Zero & $p_2$& $-0.25$&$0.06$& photometric bias\\
\hline
\end{tabular}
\caption{The mean and $1\sigma$ error of the parameter constraints from CFHTLenS using the 3D cosmic shear method, assuming all other parameters 
are fixed at the Planck (2013) maximum likelihood values.}
\label{resultstab}
\end{table*}

\begin{figure*}
    \includegraphics[angle=0,clip=,width=2\columnwidth]{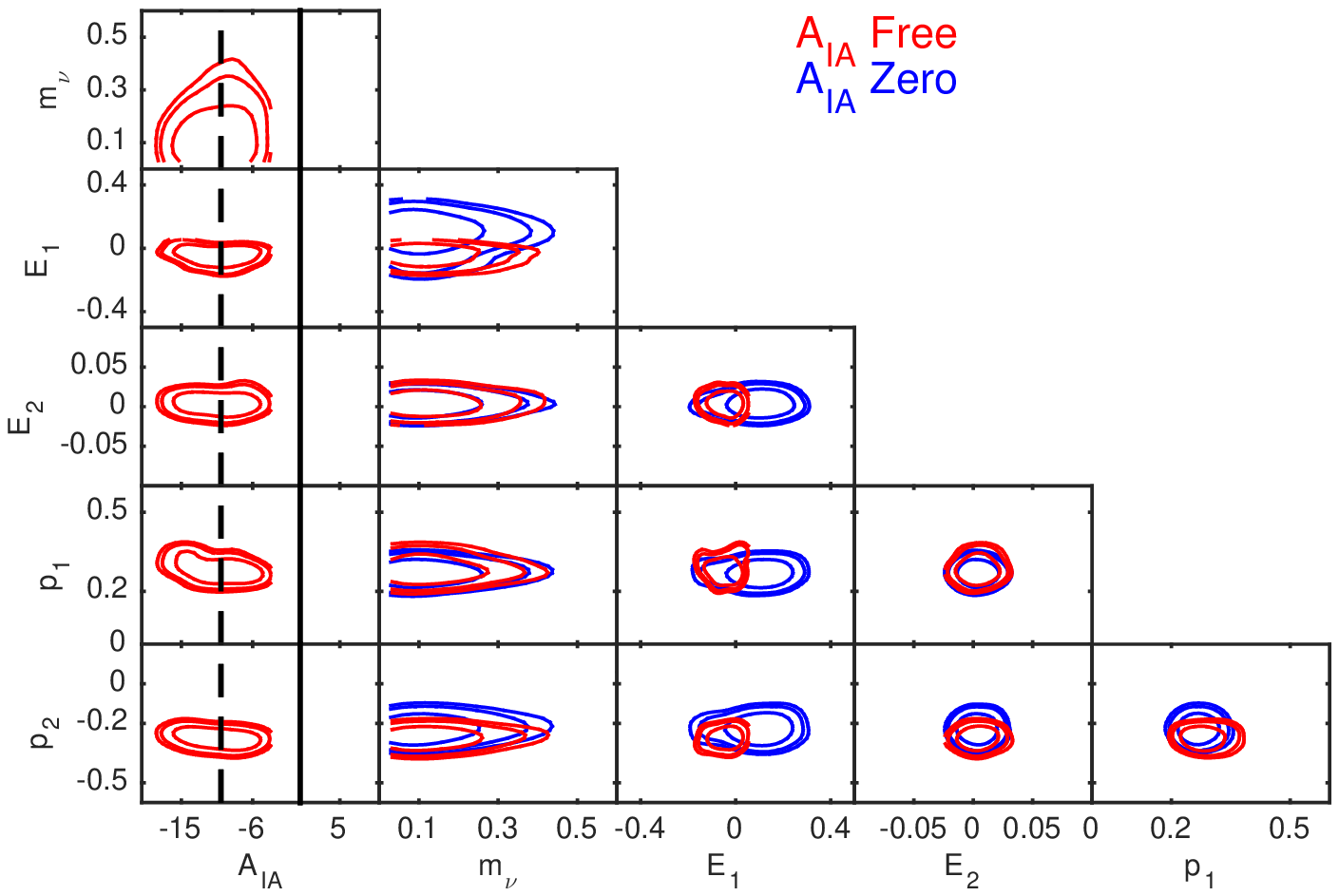}
 \caption{Projected $1$, $2$ and $3\sigma$ 
   parameter constraints for the additional parameters the intrinsic alignment amplitude $A_{\rm IA}$, the 
   sum of the neutrino masses $m_{\nu}$, the two baryonic feedback parameters $E_1$ and $E_2$, and the two photometric 
   redshift bias parameters $p_1$ and $p_2$. In this analysis we use $k_{\rm max}=5h$Mpc$^{-1}$. The red contours allow for a free 
   intrinsic alignment amplitude, and the blue contours fix its value at zero. The vertical black solid and dashed lines are 
   at $A_{\rm IA}=0$ and $-10$ respectively for reference. Cosmological parameters are set at Planck best-fit values.}
 \label{2Dproj}
\end{figure*}

\begin{figure*}
   \large {\bf $A_{\rm IA}$ zero}
    \includegraphics[angle=0,clip=,width=2\columnwidth]{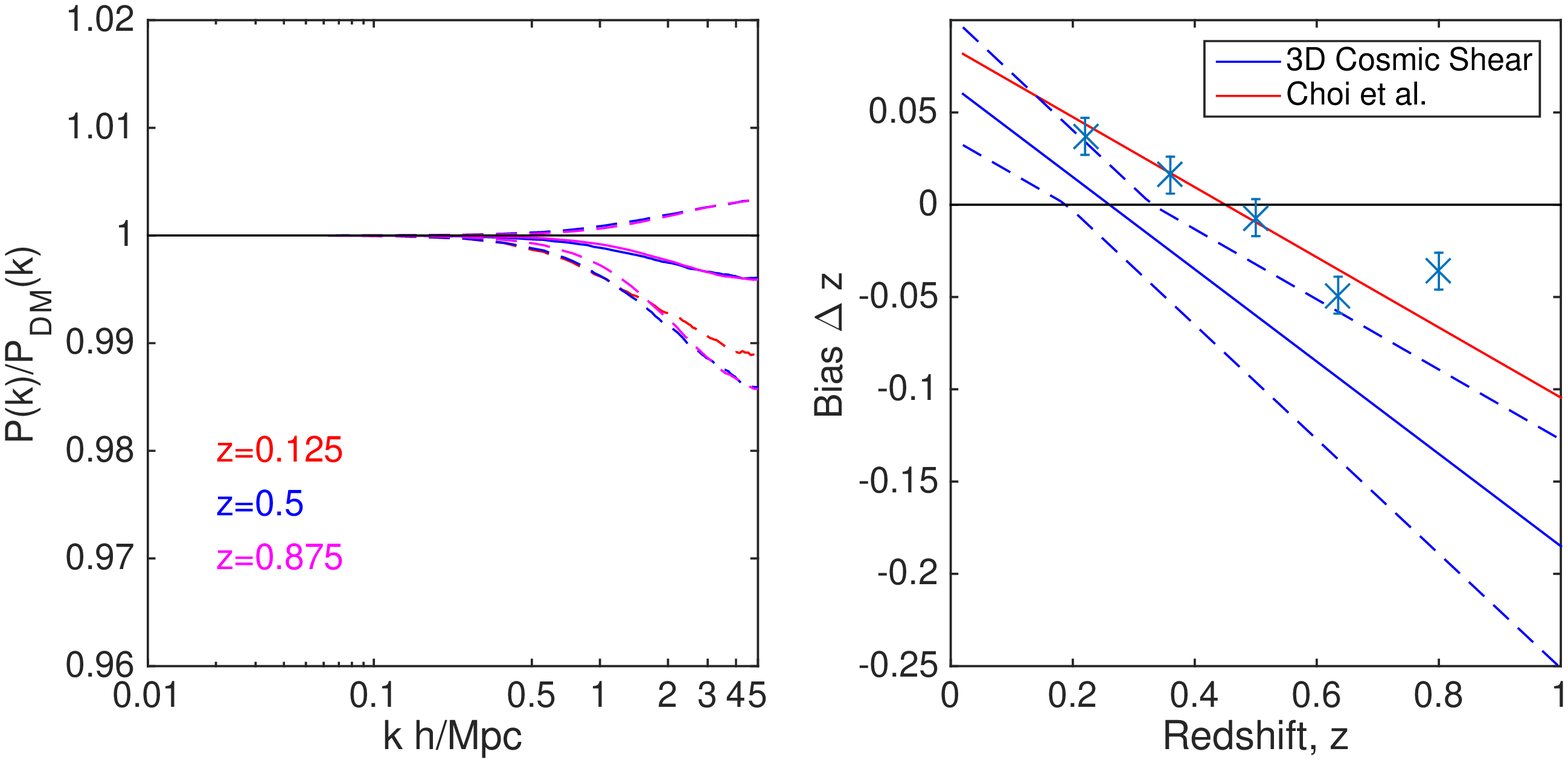}
    \\
   \large {\bf $A_{\rm IA}$ free}
    \includegraphics[angle=0,clip=,width=2\columnwidth]{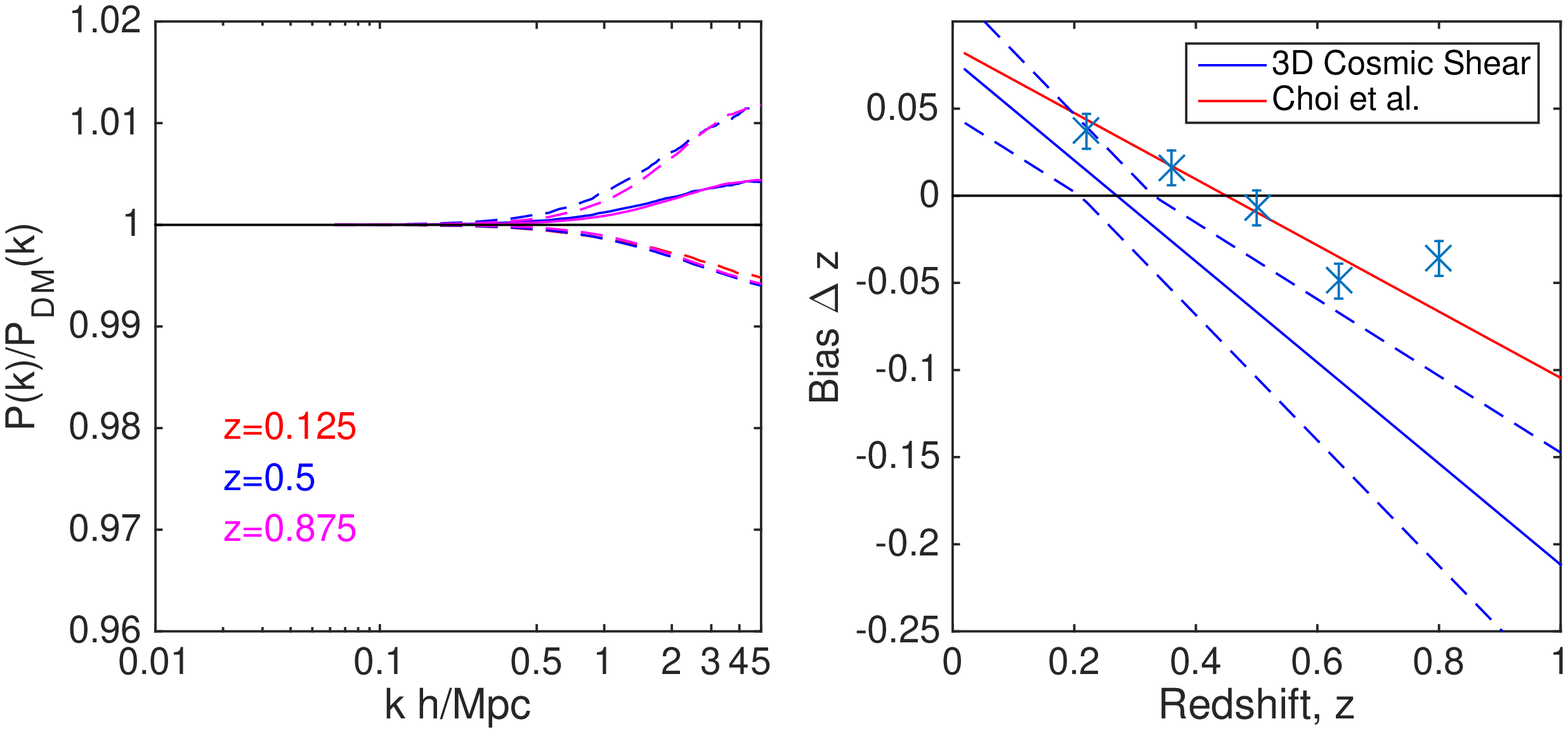}
 \caption{The best-fit functional forms for the baryonic feedback model parameters and the photometric redshift bias parameters. 
   For the baryonic feedback model we show the suppression for three representative redshifts. For the photometric 
   redshift bias we also show the Choi et al. (2015) data points (crosses) and the best fitting linear relation 
   for these (red line). The solid blue line in the righthand panel is the best fit from the cosmic shear constraints, and the 
   dashed lines are the $1\sigma$ confidence regions. 
   The legend denotes the redshifts in the lefthand panels, where the solid line are the best functions 
   and the dashed lines are the $1\sigma$ confidence regions.
   The  lower panels show the results for the case that the intrinsic alignment amplitude is a free parameter (which maximises 
   at $A_{\rm IA}\simeq -11$). The upper panels show the case in which the intrinsic alignment amplitude is set to zero.}
 \label{bestfits}
\end{figure*}

In Figure \ref{2Dproj}, and tabulated in Table \ref{resultstab}, 
we show the projected constraints on each of these parameters for two cases, one where we have left the 
intrinsic alignment amplitude to be a free parameter in the fit, and secondly where we have fixed the intrinsic alignment 
amplitude to be zero. It can be seen that the data favours a very negative intrinsic alignment amplitude parameter if allowed to. 
This is an unphysical regime for this parameter -- which should be positive if early-type galaxies are radially aligned to local 
dark matter over-densities, and cause a suppression in the cosmic shear power spectrum. 
In this analysis we also find a large photometric redshift bias,
which is consistent with, but slightly more pronounced than, the results from Choi et al. (2015)
which come from an entirely independent analysis of the photometric redshifts themselves.
We also find that a non-zero neutrino 
mass (conditional on all other $\Lambda$CDM parameters being fixed) is not favoured by the data,
with the analysis setting an upper 
limit of $m_{\nu}\ls 0.28$ eV  ($1$-$\sigma$), which 
is in agreement with other recent cosmological constraints 
(e.g Cuesta et al., 2015; Verde et al., 2014; Gonzalez-Morales et al., 2011).
In Figures \ref{2Dproj} and \ref{bestfits} we also show the case that 
the intrinsic alignment amplitude is fixed to zero. This is 
a more physical case, as there is no strong evidence for 
intrinsic alignments in the early-type galaxy sample that we use in 
our analysis (see e.g. Mandelbaum et al., 2011 and Joachimi et al., 2015). We again find 
that the neutrino mass is 
consistent with zero in this case. 
In Figure \ref{bestfits} we show the best fitting baryonic feedback parameters in this case, 
we find consistency with no baryonic feedback at all, and a tight upper limit (at 68\%) 
of 1.5\%, at $k=5h$Mpc$^{-1}$.
The amplitude of the $E_1$ parameter indicates that the mean correction must be smaller than half of that
predicted by OWLS simulations and is very far from that predicted by Illustris. This can be understood by considering equation (18). For the OWLS case  $E_1$ should be $1$ (see equation 15). For recovering the Illustris 
suppression $E_1$ should be $\sim 7$ and $E_2$ should be  small (see Figure 3);  
$E_1<1$ implies a mean correction  (i.e., a mean fractional correction to $P_{\rm DM\, only}$) smaller 
than in the  OWLS case.

Providing a $\chi^2$ goodness of fit estimate is not possible for our method, 
this is because we fit vary the covariance in the likelihood not the mean, and the variance will just 
adjust as required; therefore a Bayesian evidence calculation is required to test models correctly, but in 
order to implement such a test more code development of {\tt 3dfast} is required. In the meantime here 
we quote the likelihood values at the best-fit Planck cosmology 
for the three cases we investigate ($\Lambda$CDM, $\Lambda$CDM-fixed-$A_{\rm IA}$ free, 
and $\Lambda$CDM-fixed-$A_{\rm IA}$ zero) that are indicative of level of change in information content 
in the fits. These are $\ln(L|{\rm Planck})_{\Lambda{\rm CDM}}=17603$, 
$\ln(L|{\rm Planck})_{\Lambda{\rm CDM}-A_{\rm IA}-{\rm free}}=17607$ 
and $\ln(L|{\rm Planck})_{\Lambda{\rm CDM}-A_{\rm IA}-{\rm zero}}=17603$. These are as expected, higher for the 
$A_{\rm IA}$ case and lower for the other two cases.

The results we present are consistent with those found in 
Battye et al. (2015), MacCrann et al. (2014), Dossett et al. (2015), Joudaki 
et al. (2016) who all investigated the CFHTLenS-Planck constraints. 
Joudaki et al. (2016) most recently found the CFHTLenS data to prefer a large and negative 
intrinsic alignment amplitude, a small baryonic component,  and a 
small photometric redshift bias. A further complicating factor 
for correlation function methods is the mapping of the kernel to $k$-space, which is more complex than 
for the spherical-Bessel transform. MacCrann et al. (2014) show the kernel for a fixed redshift, and 
 Asgari et al. (2012) and Asgari \& Schneider (2015)
show that the angle-to-$\ell$ mode mapping can be complicated for a COSEBI weighting. 
A full investigation of 
the correlation function $k$-mode sensitivity is yet to be done. However, using the Bessel function relation,  
appropriate for the spherical-Bessel transform used in this paper, $\ell_{\rm max} \simeq k_{\rm max}r[z]$, 
the range of $k$-modes we probe approximately corresponds to a redshift-dependent minimum angular scale of 
$\theta_{\rm min}[z]=360/(k_{\rm max}r[z])$; which for $k_{\rm max}=1.5h$Mpc$^{-1}$ is 
$\theta_{\rm min}[z]=\{17,4,3\}$ arcminutes for $z=\{0.2,1.0,1.2\}$. On the large scales the maximum angular range is 
also affected by the Limber function assumption, which is only applicable for $\ell_{\rm min}\gs 200$ (or 
$\theta_{\rm max}\ls 100$ arcminutes; Simon, 2007, Loverde \& Afshordi, 2008), used in the theoretical interpretation of these 
papers results, which we do not assume in our analysis. 

The bias on photo-z obtained in Figure \ref{bestfits} is similar in amount and redshift dependence to 
the estimated one by Niemack et al. (2009). These authors constructed different estimators of 
photo-z for different wavelength coverage and stellar populations models. They found that lack of 
inclusion of ultra-violet filters resulted in a bias on the photo-z estimated redshift. In particular, 
comparison of their upper panel Figure 4 with our estimated photo-z bias shows a strong resemblance 
over the applicable redshift ranges.
 The photometric bias result is robust to assumptions 
about the intrinisic alignments. In all cases we find that the neutrino mass constraints 
are unchanged, and the baryonic feedback model is consistent with zero. Furthermore 
as shown in Kitching et al. (2014) the ability of the CFHTLenS data to constrain any more than a 
handful of parameters is limited. Therefore simultaneously varying 
LCDM parameters and photometric redshift bias would result in very broad, 
and inconclusive, parameter constraints that we do not show in this paper.

For comparison with other surveys' cosmic shear results we note that 
other recent cosmic shear results are not in tension with Planck. 
In particular the Deep Lens Survey (DLS; Jee et al., 2013, 2015), and the 
Dark Energy Survey (DES; The Dark Energy Survey Collaboration et al., 2015) both find cosmic shear results 
(using correlation function methodology for DLS, and correlation function and band-power methods for DES) that are 
consistent with the Planck results. An alternative explanation for the CFHTLenS descrepancy is discussed in Liu et al. (2016) 
who claim that a residual magnitude-dependent mutliplicative bias can alleviate the tension.

 It is interesting to note that of all the models we investigate for baryonic feedback 
in this paper, only OWLS AGN reproduces the gas fractions 
inferred from X-ray observations of clusters (see e.g. McCarthy et al., 2010). 
The other OWLS gas fractions are too high, 
while Illustris gas fractions are too low. As shown in Semboloni et al. (2011) using a halo model, 
the gas fractions are likely to determine the large-scale effect on the power spectrum; 
the smaller the gas fraction, the greater the suppression of the power spectrum on large 
scales (Schaye, private communication).

Therefore a suppression much smaller than that seen in the OWLS AGN output 
may be hard to reconcile with the X-ray observations of clusters. This means that the real 
tension may now be between cosmic shear and Planck constraints, and those from X-ray observations.

\section{Conclusions}
\label{Conclusion}
In this paper we present constraints using 3D cosmic shear, 
where the 3D power spectrum of weak lensing data 
is used to perform cosmological parameter inference. We improve this 
method over previous implementations by  
increasing the wavenumber resolution by a factor of $10$. We also test this method, in particular the pseudo-$C_\ell$ aspect 
that accounted for survey masks, by applying the method to the CFHT CLONE simulations. We demonstrate that we recover the input 
cosmology of these simulations that have a realistic mask, and galaxy properties similar to the CFHTLenS data. 
We then apply this method to the CFHTLenS data, as was done in Kitching et al. (2014) and recover the result of that paper: 
that on linear scales $k\leq 1.5h$Mpc$^{-1}$ the constraints are consistent with the Planck parameter constraints, but 
that on non-linear scales of $k\leq 5h$Mpc$^{-1}$ there is a mild tension with the Planck data in the $(\sigma_8$, 
$\Omega_{\rm M})$ plane. 

To investigate this tension we extend the cosmological modelling in four ways, 
each of which may account for an apparent drop in power at high-$k$, 
compared to $\Lambda$CDM. Firstly we develop a model-agnostic baryonic 
feedback approach and apply this to the OWLS and Illustris simulations. This extracts the impact of baryonic feedback on the 
matter power spectrum using a PCA method; this is complementary to more analytic physically-motivated models (such as 
those presented in Semboloni et al., 2011, Fedeli 2014, and Mead et al., 2015) but that are not guaranteed to capture all 
behaviour efficiently from simulations. This results in two additional parameters that describe potential matter power spectrum 
suppression as a function of redshift and scale. The second way we extend the method is to include intrinsic alignment 
modelling. For this we use the linear alignment model of Hirata \& Seljak (2004) with the ansatz 
of using the non-linear power spectrum. 
Thirdly we include a possible redshift-dependent photometric redshift bias. For this we 
use a linear form to minimise the number of free parameters, resulting in two additional parameters; however any functional form 
or binning in redshift could be used. Finally we include neutrino mass as an additional cosmological parameter. 

We apply 3D cosmic shear to the CFHTLenS data varying the additional parameters. With the caveat 
that for computational reasons we keep  all other 
parameters fixed at the Planck best-fit values (although this is unlikely to be a significant 
issue since Planck errors are much smaller than those from 
CFHTLenS). We find that when the intrinsic alignment amplitude is allowed to vary 
as a free parameter the data favours a large and 
negative value. This is probably unphysical:  the intrinsic alignment function is being used to boost the 
cosmic shear power, rather than suppress it as expected if tidal effects align galaxies radially near mass concentrations. In this case we also find a negligible suppression of the matter power due to baryonic feedback modelling, 
a large photometric redshift bias, and a small neutrino mass $\ls 0.28$ eV. 
If we restrict the intrinsic alignment amplitude to be zero, which is consistent with 
galaxy-galaxy lensing measurements for 
the early-type galaxy sample we use in our analysis (see Mandelbaum et al., 2011; Joachimi et al., 2015), 
then we also find that the data favours a model in which there is 
 little or no suppression of power caused by baryonic feedback effects 
and a large photometric redshift bias. 

Conditional on the Planck best fit cosmology, and further unaccounted for systematics in the CFHTLenS data, 
these results rule out the baryonic feedback models in OWLS with AGN and Illustris simulations at high significance. We find this result is robust to the amplitude of the 
intrinsic alignment signal and neutrino mass. 
To summarise: assuming Planck best-fit cosmological parameters, our 3D weak lensing analysis of CFHTLenS weak lensing data shows no evidence for either non-zero neutrino masses or baryon feedback. For physically reasonable intrinsic alignments, the data indicate a significant bias in the CFHTLenS photometric redshifts, which is very similar to, and consistent with, findings of Choi et al. (2015) based on an entirely different argument from comparison with spectroscopic samples.  When this bias is accounted for, the evidence for baryon feedback goes away.

In assessing cosmological large-scale-structure statistics, the critical 
methodological factor is the ability of methods to probe cleanly 
defined ranges of physical scales in the analysis. This is particularly crucial in cosmic shear analyses where several 
poorly understood systematic and astrophysical effects can have a large impact, and where there is potentially a wealth of 
cosmological information. The 3D cosmic shear approach taken in this paper can separate scales in this manner, and in 
addition works in the correct geometry for the data. Future optimisation of this approach will improve these aspects further 
allowing for robust scale-dependent tests of cosmology and astrophysics and, as the  volume 
of weak-lensing surveys increases in size (O(1000) sq. deg.) and depth significantly beyond the CFHTLenS data, we envision that a clear signature of 
neutrino physics will be unveiled in the sky (Jimenez et al., 2010). 

\section*{Acknowledgments}
LV thanks M. Vogelsberger for providing tabulated form of Figure 5 from  Nature 509, 177 (2014).
TDK is supported by a Royal Society University Research Fellowship.  
LV and RJ  acknowledge support of Mineco grant AYA2014-58747-P,  MDM-2014-0369 of ICCUB (Unidad de Excelencia `Maria de Maeztu') and  Royal Society grant IE140357.
We thank the CFHTLenS, Illustris, OWLS and Planck groups for making their data, catalogues and 
simulations public. We thank Gabriel Perez and the system administration of the Hipatia computer. We thank Justin Alsing and 
Jason McEwen for useful discussions.  We thank Joop Schaye, Hendrik Hildebrandt, Catherine 
Heymans, Peter Schneider, Aurel Schneider, Marika Asgari, Jia Liu, James Jee for comments on the text. 
We also thank an anonymous referee for careful review of our submission to MNRAS.

\end{document}